\documentclass[english]{article}
\usepackage{mathpazo}
\usepackage[T1]{fontenc}
\usepackage[latin9]{inputenc}
\usepackage{geometry}
\usepackage{babel}
\usepackage{graphicx}
\usepackage{setspace}
\usepackage[numbers]{natbib}
\usepackage[unicode=true]
 {hyperref}
\usepackage{breakurl}

\makeatletter
\newcommand{\lyxaddress}[1]{
\par {\raggedright #1
\vspace{1.4em}
\noindent\par}
}

\makeatother

\begin{document}

\title{Global temperatures and sunspot numbers. Are they related? Yes, but
non linearly. A reply to Gil-Alana et al. (2014)}

\author{Nicola Scafetta$^{1,2}$}

\maketitle

\lyxaddress{$^{1}$Active Cavity Radiometer Irradiance Monitor (ACRIM) Lab, Coronado,
CA 92118, USA}

\lyxaddress{$^{2}$Duke University, Durham, NC 27708, USA}
\begin{abstract}
Recently Gil-Alana et al. (Physica A: Statistical Mechanics and its
Applications 396, 42-50, 2014) compared the sunspot number record
and the temperature record and found that they differ: the sunspot
number record is characterized by a dominant 11-year cycle while the
temperature record appears to be characterized by a \textit{``singularity''}
or \textit{``pole''} in the spectral density function at the \textit{``zero''}
frequency. Consequently, they claimed that the two records are characterized
by substantially different statistical fractional models and rejected
the hypothesis that sun influences significantly global temperatures.
I will show that: (1) the ``singularity'' or ``pole'' in the spectral
density function of the global surface temperature at the ``zero''
frequency does not exist - the observed pattern derives from the post
1880 warming trend of the temperature signal and is a typical misinterpretation
that discrete power spectra of non-stationary signals can suggest;
(2) appropriate continuous periodograms clarify the issue and also
show a signature of the 11-year solar cycle (amplitude $\leq0.1$
$^{0}C$), which since 1850 has an average period of about 10.4 year,
and of many other natural oscillations; (3) the solar signature in
the surface temperature record can be recognized only using specific
techniques of analysis that take into account non-linearity and filtering
of the multiple climate change contributions; (4) the post 1880-year
temperature warming trend cannot be compared or studied against the
sunspot record and its 11-year cycle, but requires solar proxy models
showing short and long scale oscillations plus the contribution of
anthropogenic forcings, as done in the literature. Multiple evidences
suggest that global temperatures and sunspot numbers are quite related
to each other at multiple time scales. Thus, they are characterized
by cyclical fractional models. However, solar and climatic indexes
are related to each other through complex and non-linear processes.
Finally, I show that the prediction of a semi-empirical model for
the global temperature based on astronomical oscillations and anthropogenic
forcing proposed by Scafetta since 2009 has up to date been successful.\\-\\
Cite as: Scafetta, N., 2014. Global temperatures and sunspot numbers. Are they related? Yes, but non linearly. A reply to Gil-Alana et al. (2014). Physica A: Statistical Mechanics and its Applications 413, 329-342. DOI: 10.1016/j.physa.2014.06.047
\end{abstract}

\section{Introduction}

\citet{Gil-Alana} (herein referred to as GYS2014) compared patterns
between a global surface temperature record and the sunspot number
record. The aim was to determined whether the two records could be
statistically related and could be described by a common fractional
model. GYS2014 could not find a clear relationship between the two
records. Consequently, GYS2014 questioned the hypothesis of a sun\textendash climate
complex coupling proposed in several papers authored by Scafetta and
collaborators (GYS2014 references \citet{Scafetta2,Scafetta1,Scafetta3,Scafetta4,Scafetta5,Scafetta6,Scafetta7,Scafetta8,Scafetta9,Scafetta10})
and by numerous other authors (GYS2014 references \citet{Douglass,Eddy,Eichler,Friis-Christensen,Hoyt,Kirkby,Lean1,Loon,Shaviv,Soon,White}).
Herein I will show that GYS2014's interpretation is incorrect because
it is based on a misunderstanding of the critiqued studies and also
on mathematical and physical misinterpretations.

The topic addressed by GYS2014 is important because it refers to the
problem of whether the temperature signal presents a complex signature
of astronomical forcings that are mostly regulated by harmonics or,
alternatively, the temperature signal is essentially an autocorrelation
fractional signal (for example a AR(1) process) regulated by internal
chaotic dynamics alone.

To do this, GYS2014 analyzed and compared the stochastic memory (self-similarity)
of the annual sunspot number and the land-ocean temperature record
using a auto-covariance like-function and the spectral density function
looking for spectral peaks at some common frequencies. They assumed
for the two time series two different fractional process-models because,
while the sunspot numbers show a strong cyclical pattern at about
the 11-year periodicity, global temperatures mainly show an increase
across the sample (period 1880-2010) with superposed some \textquotedbl{}apparently\textquotedbl{}
minor modulations. They concluded: \textit{``Due fundamentally to
the different stochastic nature of the two series we reject the hypothesis
of a long term equilibrium relationship between the two variables.
Finally, using sunspot numbers exogenously, our results reject the
hypothesis that they affect global temperatures.}''

However, the fact that temperature variations and sunspot numbers
may show different stochastic behaviors according to a specific stochastic
measure is not a novelty. The finding simply means that these two
records correspond to different physical processes, as well known.
Temperatures are the result of the action, or the reaction, of a complex
combination of terrestrial factors that can be forced by numerous
factors including specific solar-astronomical components. Sunspot
numbers depend, but not linearly, on the magnetic activity of the
Sun, on the changing geometry of its magnetic field and other things.
The two observables may be related, but for evident physical considerations
the relation is unlikely linear nor simple.

Thus, simply showing that a specific statistical analysis (namely
GYS2014's testing of the univariate statistical properties of the
two series) is unable to clearly highlight the existence of a relationship
between two physical processes does not imply that the two processes
are physically or statistically unrelated. In fact, the existence
of a statistical relationship between solar and climate records has
been established in numerous studies using alternative and advanced
pattern recognition methodologies (direct filtering comparisons, nonlinear
process comparisons, physical modeling, etc.). Thus, GYS2014 should
have demonstrated that their analysis methodology is more appropriate
than those used in the literature. However, they do not discuss this
issue leaving open the possibility that it is GYS2014's statistical
methodology and/or their interpretation that are inappropriate or
defective in some way.

I will show that GYS2014's argument is based on some misunderstandings.
In fact, none of the studies supporting the existence of a significant
link between solar-astronomical factors and climate, have claimed
that the sunspot number record mirrors the global surface temperature
record in the way GYS2014 used these records. Indeed, any direct linear
comparison between the two records could only highlight the evident
fact that the two records present substantially different patterns,
as also GYS2014 show in their figure 2. On the contrary, the sun\textendash climate
link has been often defined as ``complex'' because involves multiple
subtle mathematical and physical issues not discussed in GYS2014,
but taken into account in the scientific literature that they have
referenced.

In the following I will try to clarify the main misunderstandings
encountered in GYS2014.

\section{Understanding the periodogram of the global surface temperature record:
discrete versus continuous algorithms}

Here I will comment on the Fourier analysis proposed in GYS2014 and
on their claim that a \textit{``singularity''} or \textit{``pole}''
exists in the temperature spectral density function at the\textit{
``zero''} frequency that, being not found in the cyclical sunspot
number record, would question the existence of a significant relationship
between the two records.

Figure 1 reproduces the sunspot number record (labeled as SN in GYS2014)
and the GISS global surface temperature (labeled as GT in GYS2014)
and their periodogram panels depicted in figure 1 of GYS2014. My additions
and comments are reported in red. In particular, I had to guess and
add the labels that were missing. GYS2014 used annually sampled records.
The temperature units are expressed in hundredth of degree Kelvin
from 1880 to 2009. In the caption of their figure 1 GYG2014 write:
\textit{``In the periodograms, the horizontal axis refers to the
discrete Fourier frequencies $\lambda_{j}=2\pi j/T$ , $j=1,...,T/2$.''}
However, from the reported numbers it can only be inferred that the
index is not the frequency $\lambda_{j}$ but the Fourier frequency
number $j$ that, for a 130-year sequence, is made to vary from 1
to 65.

\begin{figure}[!t]
\centering{}\includegraphics[width=1\textwidth]{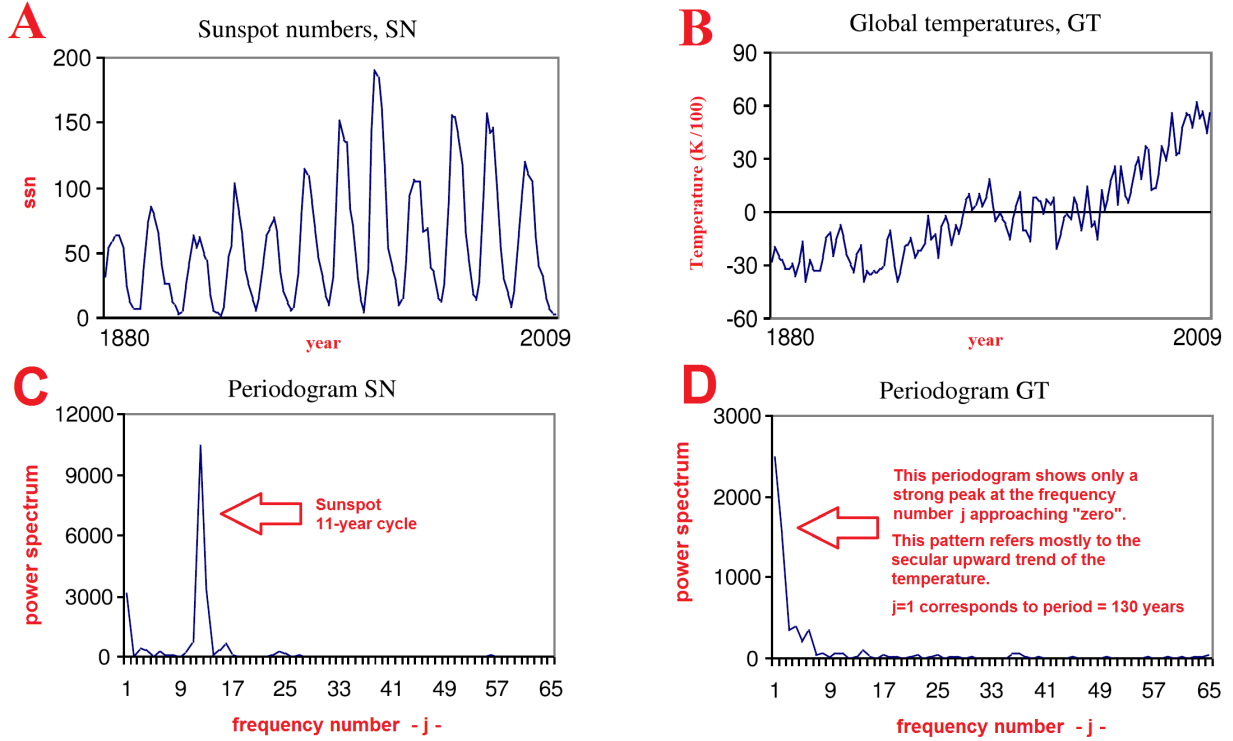}\protect\caption{Reproduction of {[}A{]} sunspot number record (SN), {[}B{]} the global
surface temperature (GT) and {[}C,D{]} their periodogram panels as
depicted in figure 1 of GYS2014. The red labels were missing in the
original figure and are guessed. Note that the periodogram is sampled
at discrete values: $j=1,2,...,65.$}
\end{figure}

GYS2014 found that the sunspot number presents a spectral peak at
about 11-year period (frequency number $j\approx12$), while for the
temperature it is stated that \textit{``the highest value in the
periodogram takes place at the smallest (zero) frequency.''} From
this evidence they concluded that while the sunspot number record
could be described by a cyclical fractional model, the global temperature
record could be described by a fractionally integrated or $I(d)$
model. Because the two stochastic models are substantially different,
GYS2014 rejected the hypothesis of a relationship between the two
variables in the long run.

A spectral pole at the \textit{``zero''} frequency is, however,
questionable given the fact that frequency numbers $0<j<1$ correspond
to periods larger than the length of the record, where spectral analysis
is not reliable any more. GYS2014 inferred a pole in the temperature
spectral density function ($f(\lambda)\rightarrow\infty$ as $\lambda\rightarrow0^{+}$)
because Figure 1D shows a strong spectral maximum of about 2500 units
at the Fourier frequency number $j=1$. However, this maximum has
a very simple explanation exposed in Figure 2 and cannot be extrapolated
to imply an infinite pole as $\lambda\rightarrow0^{+}$ and, therefore,
a fractionally integrated or $I(d)$ model does nor describe the temperature
signal.

Figure 2A shows the GISS temperature studied in GYS2014. Figure 2B
shows a high resolution \textit{continuous} periodogram of the GISS
temperature plotted against the continuous frequency expressed in
the Fourier frequency number $j$ in log-scale from $j=0.1$ to $j=65$.
The limit $j=1$, which corresponds to the lowest frequency in the
discrete Fourier transform, is highlighted with a green vertical line.

\begin{figure}[!t]
\begin{centering}
\includegraphics[scale=0.6]{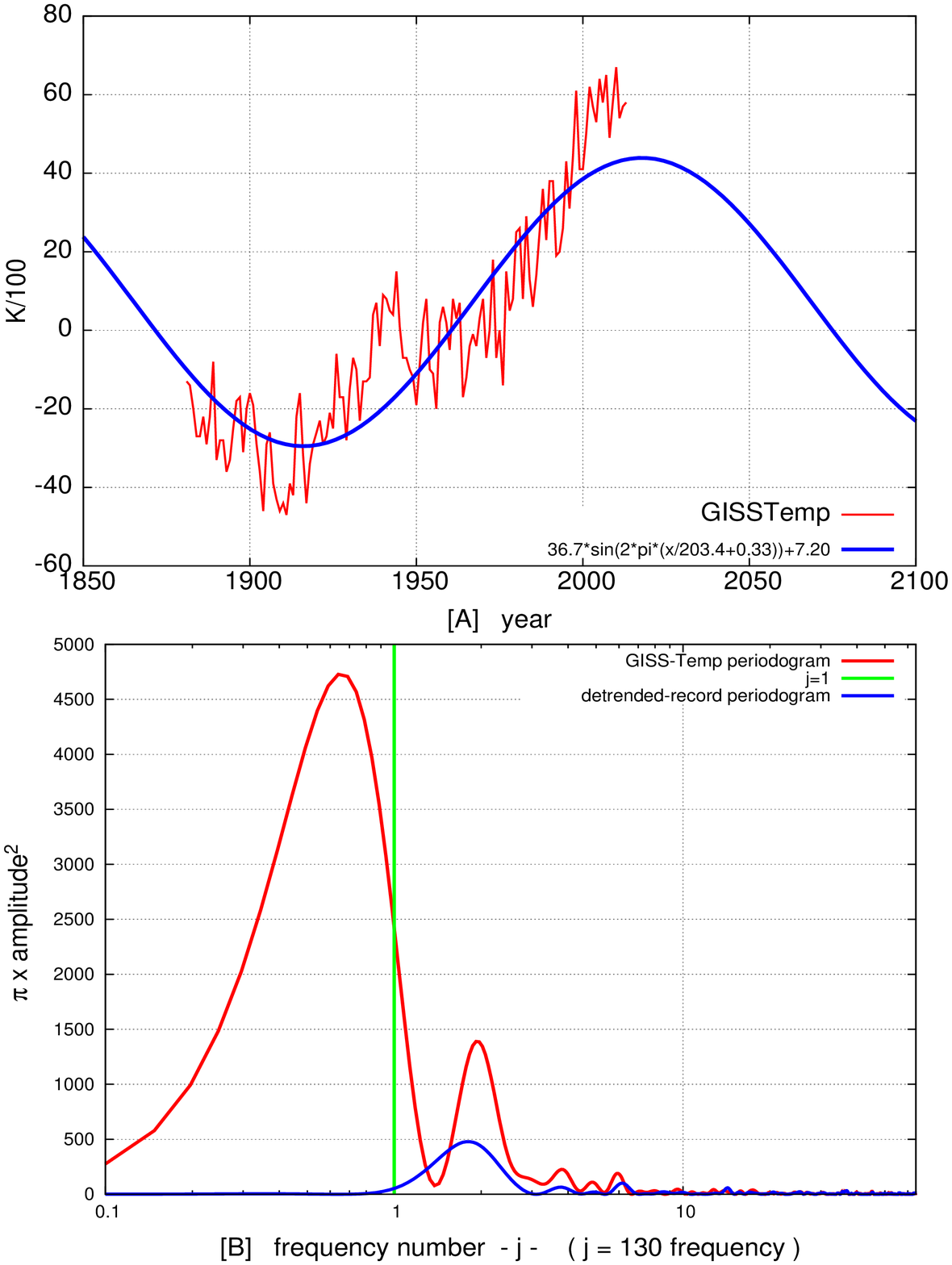}\protect\caption{{[}A{]} GISS temperature (red); Eq. \ref{eq:1} (blue curve). {[}B{]}
High resolution continuous periodogram of the original temperature
record (red curve) and of the temperature record detrended of the
harmonic given by Eq. \ref{eq:1} (blue curve). In the latter case,
GYS2014's ``singularity'' or ``pole'' in the spectral density
function at the ``zero'' frequency disappears. GISS-Temperature
data from \protect\href{http://data.giss.nasa.gov/gistemp/ }{http://data.giss.nasa.gov/gistemp/ }.}

\par\end{centering}

\end{figure}

Figure 2B shows that the temperature continuous periodogram (red curve)
intersects the $j=1$ limit at a power value of about 2500 in agreement
with GYS2014 finding of Figure 1D. However, as $j$ decreases the
power spectrum reaches a maximum at $j=0.66$ and then it converges
to zero for $j\rightarrow0^{+}$. In fact, as the frequency approaches
zero the periodogram fit curve approaches an horizontal straight line
in the sequence interval $T$. The spurious harmonic at $j=0.66$
can be calculate by regression and is given by:

\begin{equation}
h(t)=36.7\sin\left[2\pi\left(\frac{t}{203.4}+0.33\right)\right]\label{eq:1}
\end{equation}
Eq. \ref{eq:1} is shown in Figure 2A in the blue curve and clearly
suggests that it represents a spurious oscillation that the periodogram
models in its attempt of interpreting the temperature warming trend
using harmonics. If this spurious harmonic is removed from the data,
the spectral peak found in the interval $0<j\leq1$ vanishes as the
periodogram of the detrended temperature shows: see the blue periodogram
curve in Figure 2B. The periodogram for the sunspot record depicted
in Figure 1C presents an equivalent artifact at $j=1$ but it is strongly
attenuated simply because in the 1880-2010 interval this record is
more stationary than the temperature record.

In addition, the red curve of Figure 2B presents a spectral peak at
about $j=2$, which corresponds to a period of about 60 years. This
spectral peak appears to be missing in GYS2014's figure 1D because
there a discrete Fourier power spectrum is depicted at the integer
values $j=1,2,3...65$ and the discrete points were connected with
lines. This gave the misleading impression that the power spectrum
of the temperature strongly increases as $j\rightarrow0^{+},$ which
gives the impression of a ``pole'' at the ``zero'' frequency.
Therefore, GYS2014 interpretation is illusive because it derives (1)
from the discrete nature of the Fourier Analysis, (2) from improperly
connecting the spectral discrete points with lines and (3) from extrapolating
the trend of the apparent resulting curve for $0<j<1$.

Therefore, GYS2014's ``singularity'' or ``pole'' in the spectral
density function at the ``zero'' frequency does not exist. The observed
pattern emerges as an artifact that discrete periodogram algorithms
produce in their attempt of interpreting a record that within a given
interval is non-stationary, that is, presents a trend.

Thus, GYS2014's periodogram comparison is somehow misleading because
it compares the 11-year sunspot cycle against the secular warming
trend of the temperature. The latter, however, is not directly related
with the sunspot number cycle but has several physical contributions
such as the anthropogenic component plus long solar and astronomical
cycles. These cycles cannot be linearly deduced from the sunspot number
record, which shows just a dominant 11-year cycle and a small secular
modulation, but need to be evaluated using appropriate solar models
\citep{Scafetta10,Friis-Christensen,Hoyt,Kirkby,Thejll2009,Schatten,Scafetta13,Soon3,Bond,Kerr,Wang,Shapiro}.

\begin{figure}[!t]
\centering{}\includegraphics[width=0.7\columnwidth]{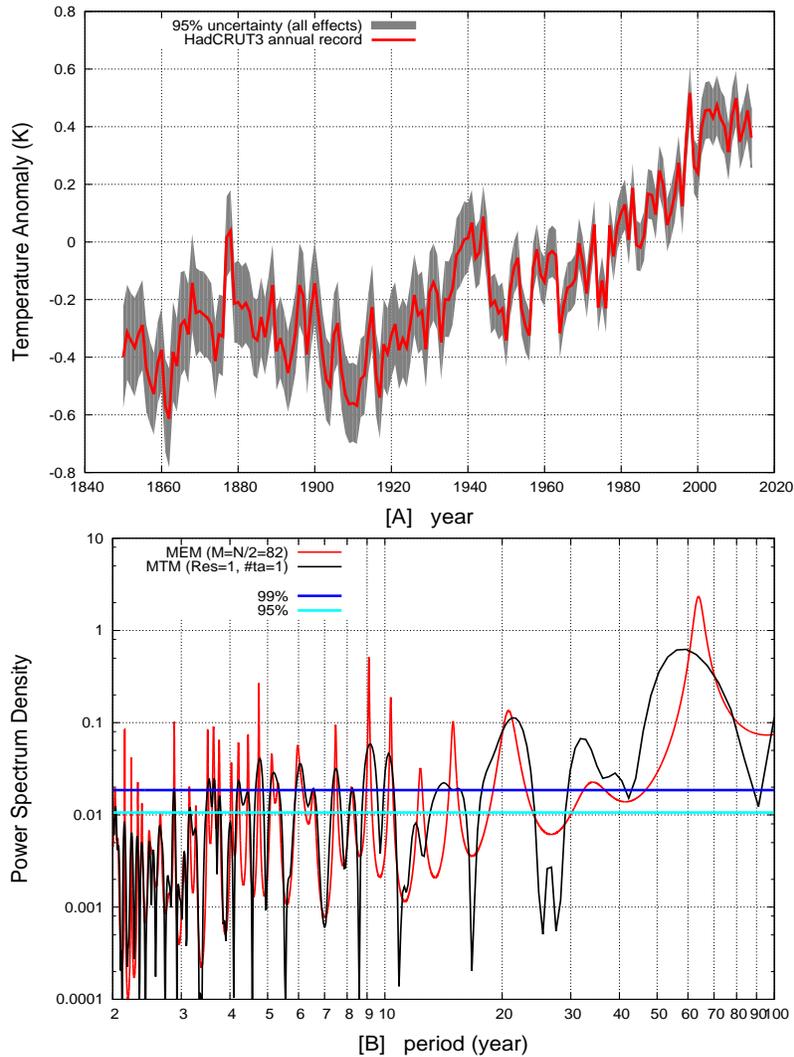}\protect\caption{{[}A{]} Annually solved HadCRUT3 global surface temperature record
\citep{Brohan2006} from 1850 to 2013. {[}B{]} Power spectrum density
functions calculated using the MEM method (using M=N/2=82) and the
MTM periodogram $f(p)$ \citep{Ghil,Press}: the calculations were
made with the SSA-MTM Toolkit. Several spectral peaks (e.g.: at about
9.1 yr, 10.4 yr, 20 yr and 60 yr) are statistically significant above
the 95\% confidence level, and their solar, lunar and astronomical
origin is explained in the literature (e.g.: \citet{Scafetta9,scafetta11,scafetta12,Scafetta13}). }
\end{figure}

To detect the 11-year solar cycle signature in the temperature record,
appropriate filtering methodologies \citep{Scafetta3,Scafetta4,Scafetta8,Scafetta10,Douglass,Lean1,White,Svensmark2007.}
or high resolution spectral techniques need to be used. Periodograms
may need to be plot in the log-log-scale, as shown below. Removing
the upward secular trend of the temperature would also be useful because
spectral techniques work better for stationary records, while a strong
upward trend introduces a significant nonstationarity that can disrupt
the spectral analysis of the signal.

\citet{scafetta11,scafetta12,Scafetta9} used advanced spectral methodologies
and showed that global surface temperature records, including the
GISS temperature used in GYS2014, present numerous specific spectral
peaks. For the benefit of the reader, Figure 3A shows the HadCRUT3
global surface temperature annual record \citep{Brohan2006}, which
was used by Scafetta in several publications where it was preferred
to the GISS temperature record used in GYS2014 because it starts in
1850. Figure 3B shows the Multi Taper Method (MTM) periodogram and
the Maximum Entropy Method (MEM) spectrum (using a pole number M=N/2,
where N=164 samples). Numerous highly confident spectral peaks are
found, and correspond to several solar, lunar and astronomical oscillations
as explained in the literature (e.g.: \citet{scafetta11,scafetta12,Scafetta9,Scafetta13}).
The statistical confidence levels (95\% and 99\%) are deduced from
the combined effects of all known physical temperature uncertainties.
The annual average range of the statistical error ($\sigma\approx0.06$
$^{o}C$) varies in time and is shown in A by the gray area. (\href{http://www.metoffice.gov.uk/hadobs/hadcrut3/diagnostics/time-series.html}{http://www.metoffice.gov.uk/hadobs/hadcrut3/diagnostics/time-series.html}):
these are given by $8.5\sigma^{2}/\pi\approx0.010$ (for the 95\%
confidence) and $15\sigma^{2}/\pi\approx0.018$ (for the 99\% confidence).

The most prominent temperature spectral peaks shown in Figure 3B are
found at these periods: $\sim$9.1 years, $\sim$10-12 years, $\sim$20
years and $\sim$60 years. The variable $\sim$10-12 year temperature
oscillation corresponds to the 11-year solar cycle that during most
of the 20th century was usually short and bounded between 10 and 11
years; its amplitude was estimated to be just a few hundredths of
degree Kelvin up to a maximum of $0.1$ K during specific periods.
Therefore, power spectra of global surface temperature records represent
the sunspot cycle as a main spectral peak at about 10.4 year period
\citep{Scafetta2012b,scafetta11,Scafetta9}. The 9.1-year cycles appears
to be related to lunar tidal cycles because it is between the 8.85-year
lunar apside rotation cycle and the second harmonic of the 18.6 lunar
nodal cycle and inclination cycle. The other spectral peaks correspond
to other solar and astronomical oscillations of the Heliosphere (e.g:
\citet{scafetta11,scafetta12,Scafetta9,Scafetta2014}). Note that
the quasi 20-year cycle may be made of three close astronomical harmonics
(see discussion in \citet[appendix]{scafetta11}): the 18.6-year lunar
cycle, the quasi 20-year oscillation induced in the Heliosphere by
Jupiter and Saturn and the quasi 22-year Hale solar magnetic cycle
that may have a high relevance for the Earth's climate, because the
Earth is imbedded in the Heliosphere.

Thus, the evidence emerging from the periodogram is that the temperature
signal is regulated by oscillations like the sunspot number record
and, therefore, belongs to the same statistical family: a cyclical
fractional model plus, eventually, a non-cyclical anthropogenic component
during the last decades.

\section{Evidences from the scientific literature on short and long scale
correlations between solar and climatic records}

About the sunspot number and the temperature records GYS2014 claimed
that \textit{``no relationship between the two variables in the long
run''} exists. In this section I will show that part of the misunderstandings
were due to the fact that GYS2014 used as solar index the sunspot
number record. GYS2014's methodology implicitly assumes that the sunspot
number record is linearly related to the other solar observables.
They also claimed that the total solar irradiance increases with the
number of sunspots because \textit{``more sunspots release energy
into the atmosphere.''} However, these arguments are physically inaccurate.

It is well-known that sunspots release \textit{less} energy than the
surrounding solar surface. During solar maxima, the total solar irradiance
(TSI) does not increase because of the energy released by the sunspots,
which is less, but mostly because of the energy released by the solar
faculae, which are hot-spots forming on the solar surface. During
solar maxima, the faculae number and area increase and their strong
\textit{brightness} overcomes the sunspot \textit{darkness}. In addition
to the irradiance contributions from faculae and sunspots, which mostly
regulate the TSI high frequency component, most of the TSI contribution
comes from the quieter extended solar regions. The emission from the
quiet regions can gradually evolve in time and regulate the TSI low
frequency component. Numerous other solar phenomena occur such as
variations of solar magnetic fields and of solar wind fluxes (\href{http://www.solarviews.com/eng/sun.htm}{http://www.solarviews.com/eng/sun.htm}).
The intrinsic complexity of the solar dynamics and of its multiple
coupled phenomena demonstrates that the sunspot number record is not
linearly related to other solar observables. Thus, the sunspot number
record cannot be used as a representative of the global solar influence
on the Earth's climate. Moreover, the climate of the Earth is regulated
by several anthropogenic and natural factors, which need to be factorized
to properly identify the solar contribution.

Let us summarize how the scientific literature cited in GYS2014 have
addressed the above issues that are necessary to understand the solar
contribution to climate change and show how closely solar and global
surface temperature records appear to be related to each other at
short and long time scales once the appropriate methodologies and
solar models are adopted.

\subsection{The sunspot number record is only a partial proxy of solar activity}

The sunspot number record is used in solar physics only as one of
the several ingredients adopted to make proxy models of solar activity.
It is these solar proxy models that are then compared in various ways
against the temperature records and/or used as external forcing of
climate models. The scientific works that GYS2014 referenced did not
base their argument using directly the sunspot number record. Those
authors noted that \textit{specific} solar proxy models correlate
quite well with the temperature patters at multiple scales. It is
those results that have suggested the existence of a significant link
between solar activity and climate. Let us clarify some of the case
addressed in GYG2014.

(1) GYS2014 claimed that \citet{Friis-Christensen} \textit{``show
that sunspot numbers between 1861 and 1989 display a significant relationship
with the Northern Hemisphere (NH) mean temperatures''.} However,
those authors did something quite different. \citet{Friis-Christensen}
found that NH land temperatures have some similarity with the 11 year
smoothed sunspot record, but the land temperature record leads the
sunspot record during some periods (e.g. the temperature peaks in
the 1940s while the sunspot record peaks around 1960), and then the
latter cannot be an usable index for the solar forcing. They then
went on to investigate the relation using the length of the solar
cycle as a proxy model for solar activity, and found good correlation
with the temperature record when the solar cycle length record is
smoothed over 5 cycles (low pass filter 1-2-2-2-1). Because of this
smoothing they really investigated only longer periods of the order
of several decadades (e.g. the Gleissberg solar spectral range of
50-90 years). Thus, \citet{Friis-Christensen} used a solar proxy
model based on the temporal duration of the \textit{solar cycle length}
and its smooths. Although, the latter was deduced from the sunspot
number record, the record is characterized by multidecadal patterns
that well correlate with the temperature record (both peak in the
1940s) but poorly correlate with the sunspot number record itself.
GYS2014 correctly highlighted that the original model proposed by
\citet{Friis-Christensen} had a problem, which was actually a border
artifact due to the smoothing that could not well model the last two
decades. However, GYS2014 did not acknowledge that the solar cycle
length model was later updated \citep{Thejll,Thejll2009,Loehle} and
only since around 1990 the type of solar forcing that is described
by the solar cycle length model no longer appears to dominate the
long-term variation of the Northern hemisphere land air temperature.
The finding implies that at least 50\% of the the post 1850 global
warming could be still associated to solar variability, which is a
finding consistent with other recent studies \citep[cf.: ][]{scafetta11,scafetta12}.

(2) The good multisecular solar-climate correlation found by other
authors (\citet{Eichler,Hoyt,Schatten,Soon}) focused on multidecadal
and secular patterns and were based on advanced multi-proxy solar
models, not on the sunspot number record alone.

(3) Some of the preliminary results of \citet{Scafetta1} and \citet{Scafetta2}
were based on a proxy solar model made of the time intervals between
major solar flares: they did not use the sunspot number record. In
this case GYS2014 referenced a critical study by \citet{Rypdal}.
However, GYS2014 did not acknowledge the rebuttal by \citet{West2010}
that demonstrated the numerous severe misconceptions of their work
made in \citet{Rypdal} who confused the time interval between flares,
which was the variable used in \citet{West2010}, for the solar flare
index that is an energy intensity record.

(4) \citet{Kirkby} does not use the sunspot number record but cosmic
ray records and their proxy models. The latter models are based on
cosmogenic nuclides such as $^{10}Be$ and $^{14}C$, whose production
on the Earth is related to the solar electromagnetic activity. These
solar proxy models present an 11-year cycle similar but not identical
to the 11-year sunspot number oscillation. In addition, while the
sunspot number is characterized by similar minima close to zero with
a relatively small low frequency component, the used proxy models
present a more significant low frequency modulation because they are
not physically bounded at zero as the sunspot number record, and well
correlate with a number of climatic indexes at multiple scales.

(6) \citet{Scafetta3} based their analysis on the ACRIM composite
of total solar irradiance satellite measurement that, contrary to
the sunspot number record, presents an upward trend from 1980 to 2000,
and better correlates with the observed temperature warming during
the same period. See also \citet{Scafettaw2009,Scafettaw2014} to
better understand the issues related to the TSI records and why they
differ from the sunspot number record.

(7) \citet{Scafetta5,Scafetta8,Scafetta10} used multi-scale thermal
models of several total solar irradiance proxy records based also
on total solar irradiance satellite measurements. Again, they did
not use the sunspot number record.

(8) Finally, Scafetta \citep{scafetta11,scafetta12,Scafetta9,Scafetta13}
used alternative and more advanced empirical models for interpreting
climate changes. These new models are based on solar and astronomical
oscillations assumed to be regulating solar activity and, directly
or indirectly, the Earth's climate plus the anthropogenic and volcano
contributions. Again, the proposed solar-astronomical proxy models
present patterns that differ from the sunspot number record.

Of course there are numerous other works showing a significant correlation
between specific solar indexes and climate records (e.g.: \citet{Mufti,Soon,Soon1,Soon3}
and many others). All solar proxy models adopted in the above studies
can present some similarity with the sunspot number record but also
strong dissimilarities, in particular in the multi-decadal and secular
scales, which are necessary to properly interpreting the low-frequency
component of the temperature record.

\subsection{Climate change is not determined by solar forcing alone}

The global surface temperature patters are not determined exclusively
by solar inputs and even less by sunspots alone. On a time scale up
to the millennial one, global climate averages are mostly regulated
by volcano eruptions, anthropogenic forcings and numerous natural
oscillations, which include solar, astronomical and lunar tidal oscillations.
To avoid misleading conclusions, the various physical attributions
need to be taken into account.

In fact, while GYG2014 simplistically compared the sunspot number
against the temperature record, many authors that they reference used
multi-attribution analysis methodologies. For example, \citet{Douglass,Lean1,Loon,Scafetta8,Scafetta10}
evaluated the signature of the 11-year solar cycle on the temperature
by simultaneously filtering off the volcano signature, the anthropogenic
signature and the ENSO oscillations. These authors found that during
the period from 1980 to 2000, which experienced very large 11-year
solar oscillations, the 11-year solar cycle signature on the global
surface temperature had an amplitude (from min to max) of about 0.1
K. Indeed, the amplitude of the 11-year solar cycle signature is very
small in the global surface temperature record. However, it increases
with the altitude. For example, in the higher troposphere the amplitude
of the 11-year solar signature can be as large as $\sim$0.4 K, as
demonstrated in several studies \citep[e.g.:][]{Scafetta10,Loon,Svensmark2007.}.

On longer time scales the solar influence becomes clearer once appropriate
solar proxy models are used. This has been demonstrated by numerous
authors such as \citet{Eddy,Hoyt,Kirkby,Scafetta10,Scafetta13}. More
recently, \citet{Steinhilber} found an excellent correlation between
a 9,400-year cosmic ray proxy model of solar activity from ice cores
and tree rings and the Holocene Asian climate as determined from stalagmites
in the Dongge cave, China. \citet{scafetta11,Scafetta13} also highlighted
a multi-secular and millennial correlation between climate records
and specific solar models. See also \citet{Scafetta8,scafetta11,scafetta12}
where the multiple contributions to climate changes are empirically
analyzed. About the strong millennial oscillation common to both solar
and temperature records see also \citet{Humlum,Bond,Kerr}, where
it was argued that the millennial natural oscillation contributed
significantly to the warming observed since 1850. In fact, the temperature
experienced significant warm periods during the Roman Optimum (100
B.C. - 300 A.D.), and during the Medieval Warm Period (900-1400 A.D.)
and significant cool periods during the Dark Age (400-800 A.D.) and
the Little Ice Age (1400-1800 A.D.) \citep{Christiansen}. Therefore,
following this millennial cycle since 1800 the temperature had to
increase naturally until the 21th century.

On the contrary, using the sunspot number record as the privileged
solar index to be compared against the temperature record GYG2014
could not take into account that: (1) the solar effect on climate
is frequency related and the effects usually get greater at lower
frequencies; (2) some climatic patterns are not related to solar variations
but have a different physical origin (e.g. the post 1850 warming can
be partially attributed to anthropogenic emissions); and (3) that
in any case in the relevant solar-climate scientific literature only
specific solar proxy models are used and these present patterns more
complex than those shown just by the sunspot number record.

After having noted that not even $CO_{2}$ and other greenhouse gases,
either of natural or of anthropogenic origin, could be the cause,
let alone the primary cause, of global climate changes, \citet{Quinn}
wrote: \textit{``Evidence indicates that global warming is closely
related to a wide range of solar-terrestrial phenomenon, from the
sun's magnetic storms and fluctuating solar wind all the way to the
Earth's core motions. Changes in the Solar and Earth magnetic fields,
changes in the Earth's orientation and rotation rate, as well as the
gravitational effects associated with the relative barycenter motions
of the Earth, Sun, Moon, and other planets, all play key roles. Clear
one-to-one correspondence exists among these parameters and the Global
Temperature Anomaly on three separate time scales.''} Many of these
issues are discussed for example in the Scafetta's papers and in other
references.

Indeed, in the last paragraph of their conclusion GYG2014 acknowledged
some limits of their analysis such as the fact that \textit{``anthropogenic
factors and other climatologic variables may be included in the model,
thus avoiding the problem that may be caused by the potential omission
of relevant variables.''} However, as explained above, several of
the studies that GYG2014 referenced have already done what GYG2014
proposed at the end of their paper, and showed that a clear solar
signature in the climate records emerges at both the short and the
long scales once that all contributions are taken into account.

\subsection{Empirical studies versus climate model studies}

It is important to clarify an issue that was only briefly addressed
in GYS2014 who correctly highlighted that there is no consensus on
the solar contribution to climate change. While empirical studies,
as those briefly summarize above, have found a strong but \textit{complex}
solar signature in the climate system at multiple timescales and have
claimed that the sun contributed at least $\sim50\%$ of the post
1850 global warming, analytical climate models, such as the general
circulation models (GCMs), have predicted only a 5\% or lower solar
contribution to the warming observed during the same period (e.g.
see the \citet{IPCC}). It is important to realize that the apparent
incompatibility between the two sets of studies is due to (1) the
different philosophical approaches used to address the problem and
(2) to the current lack of scientific understanding of microscopic
physical mechanisms regulating climate change.

In the empirical studies, researchers focus on the macroscopic characteristics
revealed by the data and provide attribution interpretations using
various forms of detailed cross-correlation pattern recognition methodologies.
The empirical/holistic approach does not require the microscopic identification
of all physical mechanisms to recognize macroscopic patterns such
as cycles, which can be directly modeled.

On the contrary, the analytical GCM approach focuses on the microscopic
modeling of the individual physical mechanisms: it uses Navier-Stokes
equations, thermodynamics of phase changes of atmospheric water, detailed
radiation budget of the Earth and atmosphere and ocean dynamics, specific
radiative forcing functions as inputs of the model, etc. The GCMs
depend on very numerous internal variables and are characterized by
serious uncertainties such as those related to the cloud formation
\citep{IPCC}, which regulate the important albedo index.

It is evident that the analytical models need to be physically \textit{complete}
to be meaningful. On the contrary, there are several reasons suggesting
that the current analytical climate models are incomplete. This explains
why there is no consensus between the empirical and analytical studies
about the solar contribution to climate change. Let us briefly summarize
some of the arguments proposed in the referenced literature.

(1) The analytical models such as the CMIP5 GCMs adopted by the \citet{IPCC}
have used a solar forcing function deduced from a total solar irradiance
proxy record that shows only a very small secular variability (e.g.
\citet{Wang}), while alternative total solar irradiance proxy models
showing a far greater secular variability and different details in
the patters also exist \citep{Hoyt,Shapiro}. These alternative solar
models better correlate with the temperature patterns on multiple
scales and reconstruct a large fraction of the warming observed since
1850 \citep{scafetta11,Hoyt,Soon,Soon1,Soon3}.

(2) The analytical models still assume that solar-climate interaction
is limited to TSI forcing alone. However, other solar-climate mechanisms
likely exist although still poorly understood. For example, the climate
system may be particularly sensitive to specific radiations (e.g.
ultraviolet light) and to cosmic ray or solar wind that could significantly
modulate the cloud cover system \citep{Kirkby}.

(3) The climatic records are characterized by numerous natural oscillations
from the decadal to the millennial timescales that have been demonstrated
to be not reproduced by the analytical models, but are present in
specific solar, lunar and astronomical records \citep{Scafetta2012b,scafetta11,scafetta12,Scafetta9,Scafetta13}.
These oscillations, including the millennial cycle, stress the importance
of solar and astronomical effects on the Earth's climate \citep{scafetta11,Steinhilber}.

In general, analytical models may theoretically be considered the
best way to exploit the confirmatory analysis. However, the exploratory
analysis - which is needed in order to envisage the primary physical
drivers of phenomena - is a completely different gnoseologic concern.
One cannot substitute the crucial stage of the exploratory analysis
with any kind of complex confirmatory mathematics. Both stages are
needed and, in general, to describe a complex system usually empirical/holistic
approaches may be more satisfactory than an analytical ones. For example,
in the analytical modeling, mistakes can be easily made when the original
set of primary drivers is speculated.

\subsection{A few examples showing a significant correlation between solar-astronomical
records and temperature records in the short and long scales}

As discussed above, the scientists who studied the solar-climate relation
and reached the conclusion of a significant solar contribution to
climatic changes did not use the sunspot record because it is substantially
different from the temperature record. To properly acknowledge the
scientific efforts and the results made by numerous specialists, I
think that it is important to briefly summarize here some of their
results for the benefit of the readers.

These few examples are useful also for responding some of the issues
addressed in GYS2014, namely that the solar signature on climatic
records can be extracted and properly recognize only using appropriate
solar proxy models, appropriate filtering and multi-forcing analysis
and, for properly interpreting the low frequency component of the
signal, long sequences should be used to avoid misinterpretations.

\begin{figure}[!t]
\begin{centering}
\includegraphics[scale=0.5]{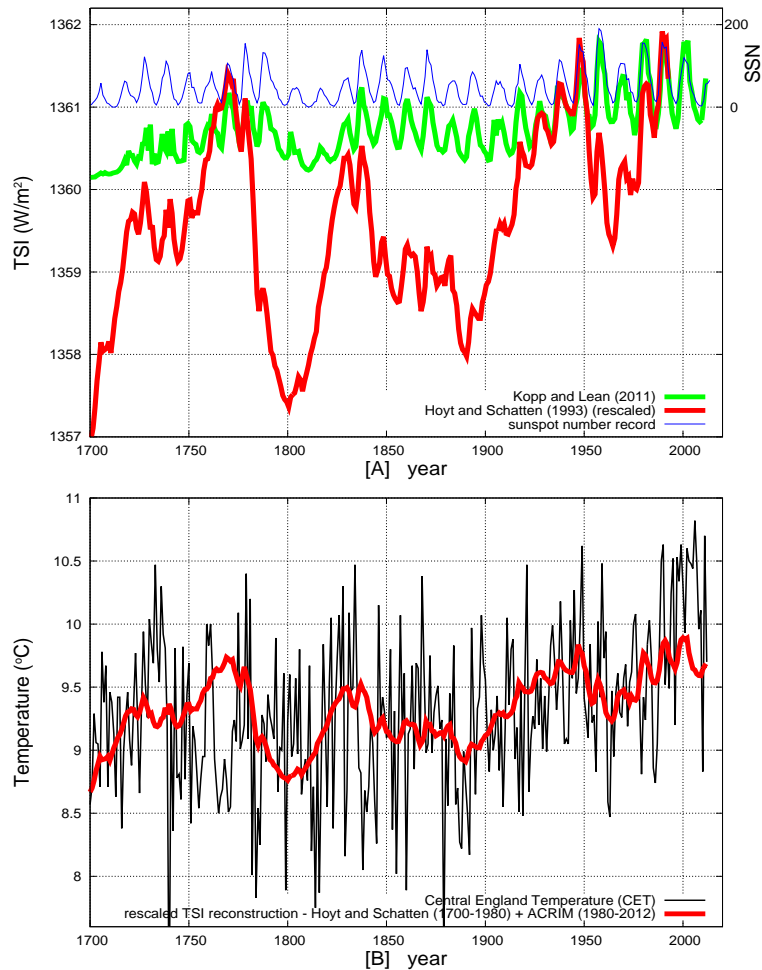}\protect\caption{{[}A{]} The sunspot number record (\protect\href{http://sidc.oma.be/}{http://sidc.oma.be/})
since 1700 versus two alternative total solar irradiance reconstructions
\citep{Wang,Hoyt}. {[}B{]} Comparison between the Central England
Temperature record \citep{Parker} and the solar reconstruction proposed
by \citet{Hoyt}: note the good correlation that includes a significant
portion of the warming observed since 1900.}

\par\end{centering}

\end{figure}

Figure 4A compares the sunspot number record since 1700 (blue curve)
versus two alternative total solar irradiance reconstructions \citep{Wang,Hoyt}.
The figure highlights that while the sunspot number is relatively
flat, solar proxy models present a more significant secular variability,
which depends greatly on the specific proxy used. Figure 4B simply
compares the Central England Temperature record \citep{Parker} and
the solar reconstruction proposed by \citet{Hoyt}. A good correlation
is noted for 300 years, which includes a significant portion of the
warming observed since 1900, which gives origin to GYS2014 putative
``zero'' frequency ``pole'', as explained in Section 2.

\begin{figure}[!t]
\centering{}\includegraphics[width=0.7\textwidth]{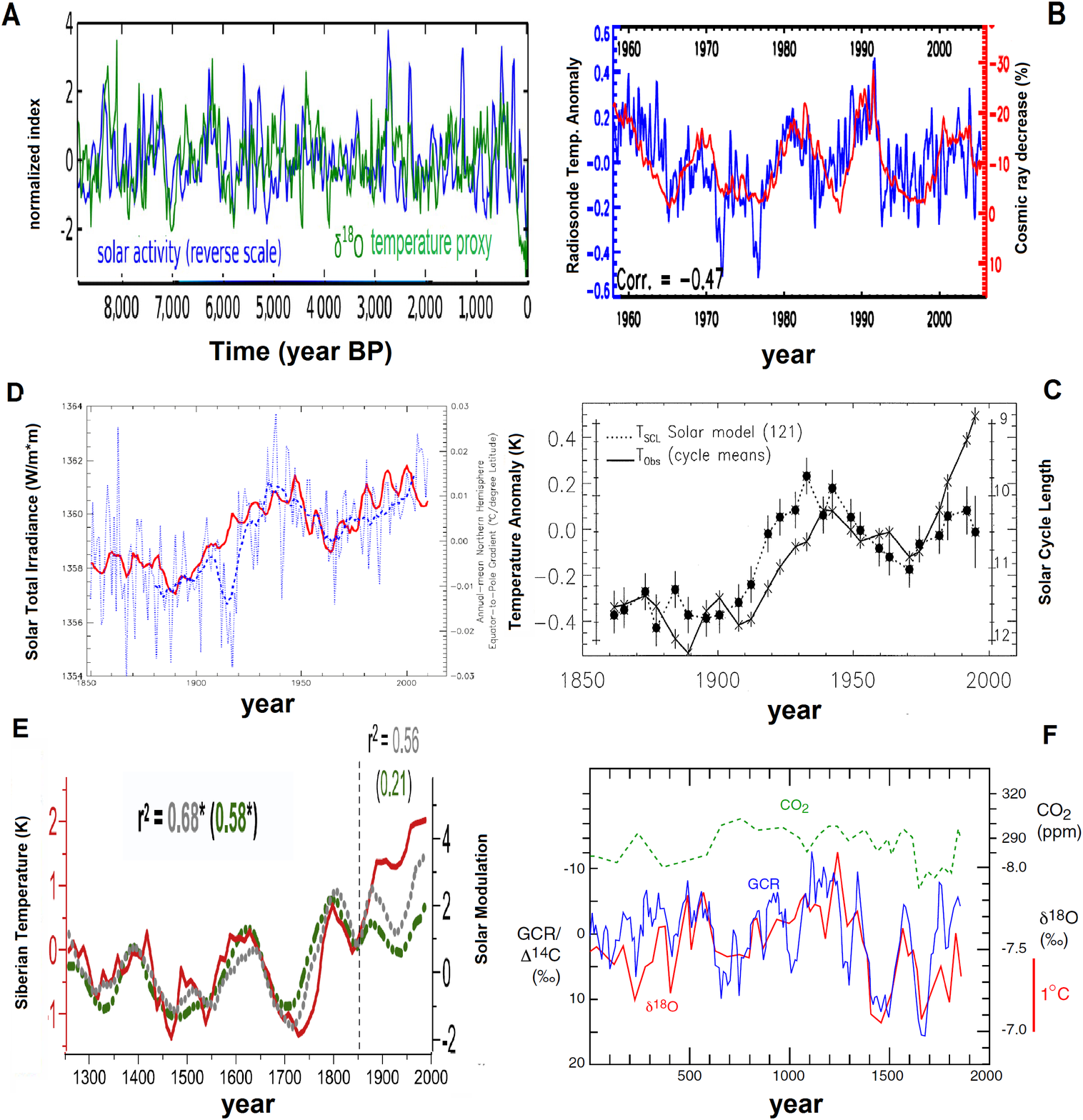}\protect\caption{{[}A{]} Comparison between a proxy of solar activity (blue) and a
proxy of temperature ($\delta^{18}O$) from Dongge cave, China, (green)
representing changes of the Asian climate during the Holocene (adapted
from \citet{Steinhilber}). The two records are evidently well correlated.
{[}B{]} Comparison between the global mean tropospheric temperatures
(blue) and the galactic cosmic ray record (red), which is modulated
by the solar magnetic activity. The panel shows the match achieved
after removing El Nino, the North Atlantic Oscillation, volcanic aerosols,
and also a linear trend from the temperature record (adapted from
\citet{Svensmark2007.}): the finding is consistent with Refs. \citep{Scafetta10,Loon}.
{[}C{]} Observed temperatures versus the SCL121 solar cycle length
model (adapted from \citet{Thejll}, cf. with \citet{Thejll2009}).
{[}D{]} Annual-mean equator to pole gradient over the entire Northern
Hemisphere (blue) and its smoothed 10-year running mean (dash blue)
versus the estimated total solar irradiance (red) of \citet{Hoyt}
(red, with up dates by \citep{Scafettaw2014}) from 1850 to 2010 (adapted
from \citet{Soon3}). {[}E{]} Comparison of the Belukha (Siberia)
temperature reconstruction with solar activity proxies (adapted from
\citet{Eichler}). {[}F{]} Temperature reconstruction for the Central
Alps over the last two millennia, obtained from the $\delta^{18}O$
composition of a speleothem from Spannagel Cave versus the variations
of cosmic rays $(\nabla^{14}C)$ and $CO_{2}$ over this period (adapted
from \citet{Kirkby}). }
\end{figure}

\begin{figure}[!t]
\centering{}\includegraphics[width=0.7\textwidth]{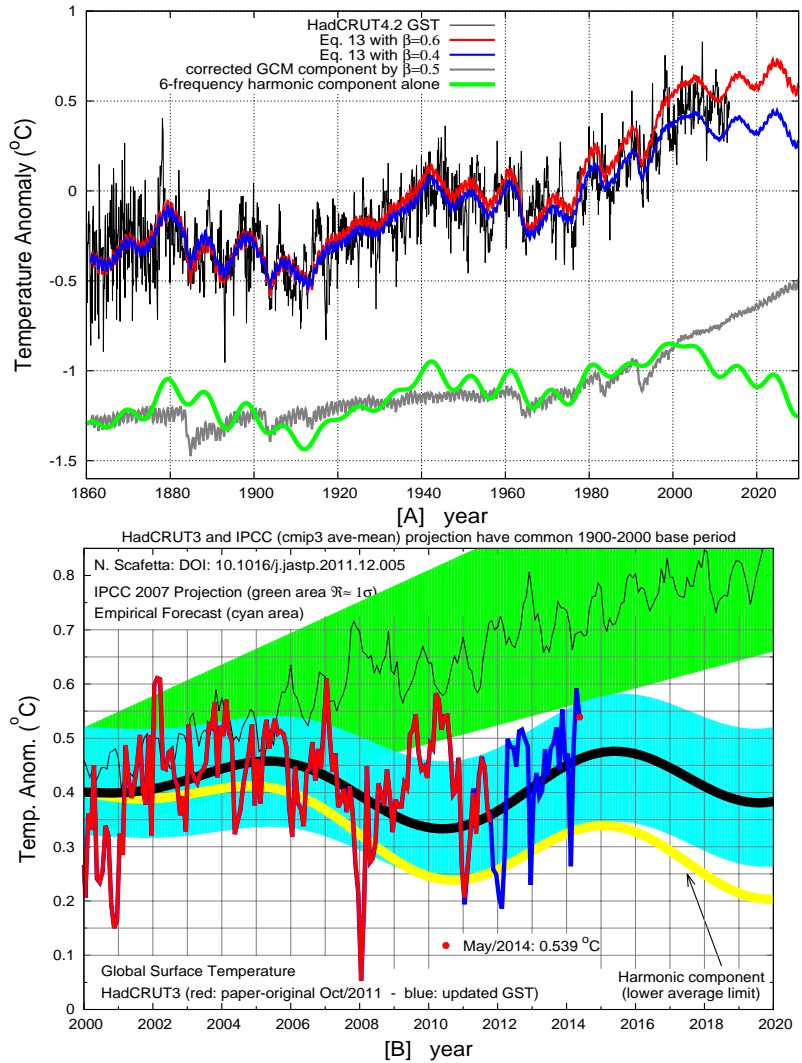}\protect\caption{Advanced modelling. {[}A{]} The semi-empirical astronomical model
for reconstructing the global surface temperature (HadCRUT4, black)
proposed by \citet[figure 25]{scafetta12}. The model (red and blue
curves) is made of two components depicted in the bottom: (1) the
gray curve is the estimate of the anthropogenic plus volcano components
made by properly attenuating the CMIP5 general circulation model ensemble
mean simulation by a factor $\beta\approx0.5$; (2) the green curve
is the estimate of the natural harmonic variability made of the 6
specific solar-astronomical harmonics from the decadal to the millennial
scale. Eq. 13 is in \citet{scafetta12} where a reader can find details
about the proposed model. {[}B{]} Detail the semi-empirical astronomical
model proposed by \citet{Scafetta2012b}. The red curve shows the
original global surface temperature record published in \citet{Scafetta2012b}.
The blue curve shows the same global surface temperature updated to
the most current available month. The back curve within the cyan area
is the semi-empirical astronomical model forecast (since 2000) that
clearly outperforms the IPCC 2007 CMIP3 general circulation model
projections (green area). The yellow curve is the harmonic component
alone without the anthropogenic component.}
\end{figure}

Figure 5 shows examples of solar-climate correlations taken from \citet{Steinhilber,Svensmark2007.,Soon3,Thejll,Eichler}
and \citet{Kirkby}. Here the correlation between solar-astronomical
records and climate records is evident at short and long time scales.
Figure 5A compares a reconstruction of solar activity and a reconstruction
of Asian climate during the Holocene (last 9000 years) \citep{Steinhilber}.
Figure 5B shows that the radiosonde temperature anomalies, after an
appropriate filtering of other climatic factors, reveals a clear signature
of the 11-year solar cycle reconstructed by the cosmic ray record
\citep{Svensmark2007.}. Figure 5C compares the instrumental global
surface temperature record versus a SCL121 solar cycle length model
\citep{Thejll}. Figure 5D compares the annual-mean equator-to-pole
gradient over the entire Northern Hemisphere versus the estimated
total solar irradiance record (red) of \citet{Hoyt} (red, with updates
by \citet{Scafettaw2014}) from 1850 to 2010 \citep{Soon3}. Figure
5E compare a Siberian temperature reconstruction with solar activity
proxies for 750 years \citep{Eichler}. Figure 5F depicts a temperature
reconstruction for the Central Alps over the last two millennia, obtained
from a $\delta^{18}O$ based temperature proxy model versus the variations
of cosmic rays $(\nabla^{14}C)$ and $CO_{2}$ over the same period
\citep{Kirkby}. All these figures suggest a strong cross-correlation
between solar proxy models and temperature patterns, which again also
include a significant fraction of the warming observed since 1850.
These empirical results indicate that the Sun has a significant influence
on the climate system, but it is not the sole contributor.

Figure 6A shows the good performance of an empirical model for the
global surface temperature proposed by \citet{scafetta11}, which
was made of 6 specific solar-astronomical cycles (periods 9.1, 10.4,
20, 60, 115, 983 yr) modeling the natural decadal-to-millennial natural
climatic oscillations plus an estimate of the anthropogenic and volcano
components made by properly attenuating the CMIP5 general circulation
model ensemble mean simulation by a factor $\beta\approx0.5$. \citet{scafetta11}
showed that his model outperforms all CMIP5 general circulation models
in reconstructing the global surface temperature record. Figure 6B
shows a detail with an update of the semi-empirical astronomical model
proposed by \citet{Scafetta2012b} in 2011. The red curve shows the
original global surface temperature record published in \citet{Scafetta2012b},
which ended in October 2011. The blue curve shows the global surface
temperature updated to the most current available month, which is
March 2014. The back curve within the cyan 1-$\sigma$ error area
is the semi-empirical astronomical model forecast (which was modeled
to start in 2000). The figure clearly shows that the proposed semi
empirical model outperforms the IPCC 2007 CMIP3 general circulation
model projections (green area) and has successfully forecast the temperature
trend from October 2011 to March 2014. Similar results are found against
the CMIP5 models \citep{scafetta11,scafetta12}. Note that a simplified
version of the same model was proposed by Scafetta since 2009 \citep{Lorenzetto,Scafetta9}
and the proposed model has correctly predicted the observed continued
standstill of the global surface temperature while the CMIP3 and CMIP5
general circulation models adopted by the \citet{IPCC} predicted
for the period 2000-2014 a strong warming of about 2 $^{o}C/century$,
which has not been observed.

Figures 4, 5 and 6 provide a strictly alternative message than figure
2 in GYS2014 that depicted directly the sunspot number versus the
temperature record in such a way to give a reader the impression that
no relationship exists between the two records.

\section{Conclusion}

GYS2014 claimed that the temperature record is characterized by a
\textit{``singularity''} or \textit{``pole''} at the \textit{``zero''}
frequency while the sunspot number record presents a dominant 11-year
cyclical pattern. From there they inferred that the two systems are
characterized by two fundamentally different stochastic natures: the
temperature signal would be a fractional integrated or I(d) model,
while the sunspot number record would be a cyclical fractional model.
Therefore, the two records were claimed to be not related.

I have demonstrated that the presumed temperature singularity at the
``zero'' frequency does not exist. GYS2014's observed pattern (a
strongly increasing power in the temperature spectrum as the frequency
number $j\rightarrow0^{+}$) is an artifact that discrete periodogram
algorithms can suggest when processing non-stationary sequences. On
the contrary, a continuous periodogram (see Figure 2B) reveals that
as the frequency approaches zero (Fourier frequency number $0\leq j<1$)
the power spectrum reaches a maximum and then it converges to zero,
not to infinity, as suggested in GYS2014. Continuous periodograms
(Figure 2B and 3B) also reveal a number of highly significant spectral
peaks suggesting that the temperature record is made of numerous oscillations
that were hidden in GYS2014 discrete periodogram.

From the continuous temperature spectra depicted in Figure 2B and
3B the only possible conclusion is that the temperature too can be
described by a cyclical fractional model like the sunspot number record,
although to properly identify the temperature harmonics proper solar/astronomical
models need to be used. This result corrects GYS2014's claim that
\textit{``Due fundamentally to the different stochastic nature of
the two series we reject the hypothesis of a long term equilibrium
relationship between the two variables.'' }As confirmed by my analysis
and explained in the literature, the temperature is likely made of
a natural harmonic components with a likely solar/astronomical origin,
plus a non-harmonic anthropogenic component particularly active during
the last decades \citep{scafetta11,scafetta12,Scafetta13}.

The good correlation between solar/astronomical models and the climate
system is obscured in GYS2014 because they used the sunspot number
record as the privileged solar index. However, the works claiming
the existence of a significant link between the sunspot and the climate
records at both the short and long time scale did not base their argument
directly on the sunspot number record. They used solar proxy models
that significantly differ from the sunspot number record in particular
in the multidecadal and multisecular modulation, in other patterns
and also for physical construction. Moreover, the referenced literature
already acknowledges and takes into account that multiple factors,
both anthropogenic and natural, modulate climatic records. These factors
need to be simultaneously factorized out to clearly reveal the solar-astronomical
component of climate changes.

In particular, the global surface temperature record contains an 11-year
solar cycle signature but, as well known, it is relatively small,
just a few hundredths of degree Kelvin, and beats with the $\sim9.1$
year soli-lunar tidal cycle. This small solar signature can be hardly
noticed without an appropriate analysis as those proposed for example
in \citet{Scafetta2012b,scafetta11,scafetta12,Scafetta9,Scafetta13}
and by numerous other authors \citep{Douglass,Hoyt,Lean1,Loon,Svensmark2007.}.
Numerous other natural oscillations linked to solar variations exist
such as cycles at 20, 60, 115, 1000 year etc. \citep{scafetta11,scafetta12,Eichler,Eddy,Kirkby,Loehle,Scafetta13,Bond,Kerr,Christiansen,Steinhilber}.
Empirical models based on these signatures have been proposed and
demonstrated to reconstruct the temperature patterns since 1850 better
than the CMIP3 and CMIP5 general circulation models adopted by the
\citet{IPCC}. In particular the solar-astronomical semi-empirical
models have correctly predicted the temperature standstill observed
since 2000 while the CMIP3 and CMIP5 general circulation models have
predicted a strong warming that has not been observed \citep{scafetta11,scafetta12,Scafetta2014}.

In conclusion, the claim that the global surface temperature record
is a fractional random signal fundamentally different from the harmonic
nature of the astronomical signals is not supported by the data and
careful analysis. The global surface temperature record appears to
be made of natural specific oscillations with a likely solar/astronomical
origin plus a non-cyclical anthropogenic contribution during the last
decades. Indeed, because the boundary condition of the climate system
is regulated also by astronomical harmonic forcings, these frequencies
need to be part of the climate signal in the same way the tidal oscillations
are regulated by soli-lunar harmonics.

\section*{Appendix}

Both MEM and the periodogram are used in Figure 3 because, as \citet[pp. 574]{Press}
wrote, \textit{\textquotedbl{}Some experts recommend the use of this
algorithm in conjunction with more conservative methods, like the
periodogram, to help choose the correct model order, and to avoid
getting too fooled by spurious spectral features.\textquotedbl{}}
Essentially, the two methodologies should be applied in the following
way: first the periodogram is used to find the spectral peaks with
their statistical confidence, then MEM is used to sharp the results.
MEM spectral peaks that are not confirmed by the periodogram should
be rejected. For sharpness the MEM order M should be chosen in function
of the spectral band that needs to be analyze: M needs to be very
high up to the maximum N/2 (N is the number of data points in the
sequence) to efficiently analyze the low frequency range \citep[supplement]{Scafetta2012b}.
Figure 2B uses MEM with $M=N/2$ and does not show spurious peaks
because each MEM peak is confirmed by a correspondent peak in the
MTM periodogram, and most peaks for 4-year and larger periods have
a 95\% and more statistical confidence relative to the physical noise
of the signal.

Power spectra are also affected by a specific resolution related to
the length $T$ of the record, which is given by $df=1/T$. Taking
into account the specific spectral resolution is important to separate
close harmonics. For example, two frequencies $f_{1}$ and $f_{2}$
can be separated only if $|f_{1}-f_{2}|\geq df$. In spectral analysis,
a spectral peak at frequency $f<2/T$ has an error of $\nabla f=\pm2/T$;
the correspondent period is $P=P_{0}\pm P_{0}^{2}/2T$. Thus, if $P_{0}$
is larger than the temporal interval of the sequence, the spectral
analysis becomes highly unstable. Moreover, to properly separate the
harmonic constituents of the temperature, if they have an astronomical
origin as suggested in \citet{Scafetta9}, spectral analysis requires
a resolution of $df\leq0.0056$ $yr^{-1}$, which corresponds to a
segment length of $L\geq1/df=178.4$ years. In fact, astronomical
frequencies are harmonics of a 178.4 year period \citep[cf. ][figure 4]{Scafetta2014}.
Temperature segments longer than 100 years would be necessary to separate
at least the major expected constituent harmonics and recognize an
astronomical influence on the climate at the decadal-secular scales.
Thus, using temperature segments of 20 to 60 years as proposed by
\citet{Holm} is inefficient to highlight a spectral coherence between
astronomical and temperature harmonics.

\newpage{}


\begin{thebibliography}{100}
\bibitem[Gil-Alana et al.(2014)]{Gil-Alana} L. A. Gil-Alana, O. S.
Yaya, O. I. Shittu. Global temperatures and sunspot numbers. Are they
related? Physica A: Statistical Mechanics and its Applications 396
(2014) 42-50.

\bibitem[Scafetta(2012b)]{Scafetta2012b} N. Scafetta. Testing an
astronomically based decadal-scale empirical harmonic climate model
versus the IPCC (2007) general circulation climate models. J. Atmos.
Sol. Terr. Phys. 80 (2012) 124\textendash 137.

\bibitem[Scafetta(2013)]{scafetta11} N. Scafetta. Discussion on climate
oscillations: CMIP5 general circulation models versus a semi-empirical
harmonic model based on astronomical cycles. Earth-Science Reviews
126 (2013) 321-357.

\bibitem[Scafetta(2013)]{scafetta12} N. Scafetta. Solar and planetary
oscillation control on climate change: hind-cast, forecast and a comparison
with the CMIP5 GCMs. Energy \& Environment 24(3-4) (2013) 455\textendash 496.

\bibitem[Lorenzetto(2009)]{Lorenzetto} Newspaper interview of N.
Scafetta by S. . \textquotedbl{}Se la Terra si surriscalda colpa del
Sole: l\textquoteright uomo non c\textquoteright entra\textquotedbl{}.
Il Giornale, October 25, 2009. \href{http://www.ilgiornale.it/news/se-terra-si-surriscalda-colpa-sole-l-uomo-non-c-entra.html}{http://www.ilgiornale.it/news/se-terra-si-surriscalda-colpa-sole-l-uomo-non-c-entra.html}

\bibitem[Scafetta et al.(2004)]{Scafetta2} N. Scafetta, P. Grigolini,
T. Imholt, J. A. Roberts, B. J. West. Solar turbulence in Earth\textquoteright s
global and regional temperature anomalies. Physical Review E 69 (2004)
026303.

\bibitem[Scafetta and West(2003)]{Scafetta1} N. Scafetta, B. J. West.
Solar flare intermittency and the Earths temperature anomalies. Physical
Review Letters 90 (2003) 248701.

\bibitem[Scafetta and West(2005)]{Scafetta3} N. Scafetta, B. J. West.
Estimated solar contribution to the global surface warming using the
ACRIM TSI satellite composite. Geophysical Research Letters 32 (2005)
L18713.

\bibitem[Scafetta and West(2006a)]{Scafetta4} N. Scafetta, B. J.
West. Reply to comments by J. Lean on estimated solar contribution
to the global surface warming using the ACRIM TSI satellite composite.
Geophysical Research Letters 33 (2006) 15.

\bibitem[Scafetta and West(2006b)]{Scafetta5} N. Scafetta, B. J.
West. Phenomenological solar signature in 400 years of reconstructed
Northern Hemisphere temperature record. Geophysical Research Letters
33 (2006) 17.

\bibitem[Scafetta and West(2007)]{Scafetta6} N. Scafetta, B. J. West.
Phenomenological reconstruction of the solar signature in the NH surface
temperature records since 1600. Journal of Geophysical Research 112
(2007) D24S03.

\bibitem[Scafetta and  West(2008)]{Scafetta7} N. Scafetta, B. J.
West. Is climate sensitive to solar variability. Physics Today 3 (2008)
50\textendash 51.

\bibitem[Scafetta(2009)]{Scafetta8} N. Scafetta. Empirical analysis
of the solar contribution to global mean air surface temperature change.
Journal of Atmospheric and Solar-Terrestrial Physics 71 (2009) 1916\textendash 1923.

\bibitem[Scafetta(2010)]{Scafetta9} N. Scafetta. Empirical evidence
for a celestial origin of the climate oscillations and its implications.
Journal of Atmospheric and Solar-Terrestrial Physics 72 (2010) 951\textendash 970.

\bibitem[Scafetta(2013)]{Scafetta10} N. Scafetta. Discussion on common
errors in analyzing sea level accelerations, Solar Trends and Global
Warming. Pattern Recogn. Phys. 1 (2013) 37-57. http://dx.doi.org/10.5194/prp-1-37-2013

\bibitem[Douglass and Clader(2002)]{Douglass} D. H. Douglass, B.
D. Clader. Climate sensitivity of the Earth to solar irradiance. Geophysical
Research Letters 29 (16) (2002) 331\textendash 334.

\bibitem[Eddy(1976)]{Eddy} J. A. Eddy. The Maunder minimum. Science
192 (1976) 1189\textendash 1202.

\bibitem[Eichler et al.(2009)]{Eichler} A. Eichler, S. Olivier, K.
Heenderson, A. Laube, J. Beer, T. Papina, H.W. Gaggeler, M. Schwikowski.
Temperature response in the Altai region lags solar forcing. Geophysical
Research Letters 36 (2009) L01808.

\bibitem[Friis-Christensen and  Lassen(1991)]{Friis-Christensen}
E. Friis-Christensen, K. Lassen. Length of the solar cycle, an indication
of solar activity closely associated with climate. Science 254 (1991)
698\textendash 700.

\bibitem[Hoyt and Schatten(1997)]{Hoyt} D. V. Hoyt, K. H. Schatten.
The Role of the Sun in the Climate Change. Oxford University Press,
New York (1997).

\bibitem[Kirkby(2007)]{Kirkby} J. Kirkby. Cosmic rays and climate.
Surveys in Geophysics 28 (2007) 333\textendash 375.

\bibitem[Lean and Rind(2009)]{Lean1} J. L. Lean, D. H. Rind. How
will Earth\textquoteright s surface temperature change in future decades?
Geophysical Research Letters 36 (2009) L15708.

\bibitem[van Loon and Labitzke(2000)]{Loon} H. van Loon, K. Labitzke.
The influence of the 11-year solar cycle on the stratosphere below
30 km. A review. Space Science Reviews 94 (2000) 259\textendash 278.

\bibitem[Shaviv(2008)]{Shaviv} N. J. Shaviv. Using the oceans as
a calorimeter to quantify the solar radiative forcing. Journal of
Geophysical Research 113 (2008) A11101.

\bibitem[Soon(2009)]{Soon} W.-H. Soon. Solar arctic mediated climate
variation on multidecadal to centennial timescales: empirical evidence,
mechanistic explanation and testable consequences. Physical Geography
30 (2) (2009) 144\textendash 148.

\bibitem[White et al.(1997)]{White} W. B. White, J. Lean, D. R. Cayan,
M. D. Dettinger. Response of global upper ocean temperature to changing
solar irradiance. Journal of Geophysical Research 102 (1997) 3255\textendash 3266.

\bibitem[Thejll and Lassen(2000)]{Thejll} P. Thejll, K. Lassen. Solar
forcing of the northern hemisphere land air temperature: new data.
J. Atmos. Solar-Terrest. Phys. 62 (2000) 1207-1213.

\bibitem[Thejll(2009)]{Thejll2009}Thejll, P., 2009. Update of the
Solar Cycle Length Curve, and the Relationship to the Global Mean
Temperature. Danish Climate Centre Report 09-01.

\bibitem[Loehle and Scafetta(2011)]{Loehle} C. Loehle, N. Scafetta,
Climate Change Attribution Using Empirical Decomposition of Climatic
Data. The Open Atmospheric Science Journal 5 (2011) 74-86.

\bibitem[Hoyt and Schatten(1993)]{Schatten} D. V. Hoyt, K. H. Schatten.
A discussion of plausible solar irradiance variations, 1700\textendash 1992.
J. Geophys. Res. 98 (1993) 18895\textendash 18906.

\bibitem[Rypdal(2010)]{Rypdal} M. Rypdal, K. Rypdal, Testing hypotheses
about sun-climate complexity linking. Physical Review Letters 104
(12) (2010) 128501.

\bibitem[Scafetta and West(2010)]{West2010}N. Scafetta, B. J. West.
Comment on `Testing hypotheses about Sun-climate complexity linking'.
Physical Review Letters 105 (2010) 219801.

\bibitem[Scafetta(2012)]{Scafetta13} N. Scafetta. Multi-scale harmonic
model for solar and climate cyclical variation throughout the Holocene
based on Jupiter-Saturn tidal frequencies plus the 11-year solar dynamo
cycle. Journal of Atmospheric and Solar-Terrestrial Physics 80 (2012)
296-311.

\bibitem[Mufti and Shah(2011)]{Mufti}S. Mufti, G. N. Shah: Solar-geomagnetic
activity influence on Earth's climate. J. Atmos. Sol. Terr. Phys.
73(13), (2011) 1607\textendash 1615.

\bibitem[Soon(2005)]{Soon1}W.-H. Soon: Variable solar irradiance
as a plausible agent for multidecadal variations in the Arctic-wide
surface air temperature record of the past 130 years. Geophysical
Research Letters 32 (2005) L16712.

\bibitem[Soon and  Legates(2013)]{Soon3}W.-H. Soon, D. L. Legates:
Solar irradiance modulation of Equator-to Pole (Arctic) temperature
gradients: Empirical evidence for climate variation on multi-decadal
timescales. J. Atmos. Sol. Terr. Phys. 93 (2013) 45-56.

\bibitem[Humlum et al.(2011)]{Humlum}O. Humlum, K. Stordahl, J.-E.
Solheim: Identifying natural contributions to late Holocene climate
change. Global and Planetary Change 79 (2011) 145-156.

\bibitem[Bond et al.(2001)]{Bond} G. Bond, B. Kromer, J. Beer, R.
Muscheler, M. N. Evans, W. Showers, S. Hoffmann, R. Lotti- Bond, I.
Hajdas, G. Bonani: Persistent solar influence on North Atlantic climate
during the Holocene. Science 294 (2001) 2130\textendash 2136.

\bibitem[Kerr(2001)]{Kerr} R. A. Kerr: A variable sun paces millennial
climate. Science 294 (2001) 1431-1433.

\bibitem[Christiansen and Ljungqvist(2012)]{Christiansen}Christiansen,
B., Ljungqvist, F.C., 2012. The extra-tropical Northern Hemisphere
temperature in the last two millennia: reconstructions of low-frequency
variability. Clim. Past 8, 765\textendash 786.

\bibitem[Svensmark and  Friis-Christensen(2007)]{Svensmark2007.}
H. Svensmark, E. Friis-Christensen. Reply to Lockwood and Fr\"ohlich
- The persistent role of the Sun in climate forcing. Danish National
Space Center Scientific Report 3/2007. Available at \href{http://www.space.dtu.dk/english/research/reports/scientific_reports}{http://www.space.dtu.dk/english/research/reports/scientific\_{}reports}

\bibitem[Scafetta and Willson(2009)]{Scafettaw2009} N. Scafetta,
R. Willson, 2009. ACRIM-gap and Total Solar Irradiance (TSI) trend
issue resolved using a surface magnetic flux TSI proxy model. Geophysical
Research Letter 36, L05701. DOI: 10.1029/2008GL036307.

\bibitem[Scafetta and Willson(2014)]{Scafettaw2014} N. Scafetta,
R. C. Willson. ACRIM total solar irradiance satellite composite validation
versus TSI proxy models. Astrophysics and Space Science (2014). DOI:
10.1007/s10509-013-1775-9.

\bibitem[Steinhilber et al.(2012)]{Steinhilber} F. Steinhilber, J.
A. Abreu, J. Beer, I. Brunner, M. Christl, H. Fischer, U. Heikkilä,
P. W. Kubik, M. Mann, K. G. McCracken, H. Miller, H. Miyahara, H.
Oerter, F. Wilhelms: 9,400 years of cosmic radiation and solar activity
from ice cores and tree rings. PNAS 109 (16) (2012) 5967\textendash 5971.

\bibitem[Quinn(2010)]{Quinn}J. M. Quinn: Global warming. Geophysical
counterpoints to the enhanced greenhouse theory. Dorrance Publishing
Co., Inc., Pittsburgh, USA (2010).

\bibitem[IPCC(2013)]{IPCC} Intergovernmental Panel on Climate Change
(IPCC). Climate Change 2013: The Physical Science Basis: Fifth Assessment
Report (2013). Available at \href{http://www.ipcc.ch/}{http://www.ipcc.ch/}

\bibitem[Wang et al.(2005)]{Wang} Y.-M. Wang, J. L. Lean, N. R. Sheeley
Jr. Modeling the sun's magnetic field and irradiance since 1713. Astrophys.
J. 625 (2005) 522\textendash 538.

\bibitem[Shapiro et al.(2011)]{Shapiro} A. I. Shapiro, W. Schmutz,
E. Rozanov, M. Schoell, M. Haberreiter, A. V. Shapiro, S. Nyeki. A
new approach to the long-termreconstruction of the solar irradiance
leads to large historical solar forcing. Astron. Astrophys. 529 (2011)
A67.

\bibitem[Brohan(2006)]{Brohan2006} Brohan, P., Kennedy, J.J., Harris,
I., Tett, S.F.B., Jones, P.D., 2006. Uncertainty estimates in regional
and global observed temperature changes: a new dataset from 1850.
Journal of Geophysical Research 111, D12106. doi:10.1029/ 2005JD006548.

\bibitem[Ghil et al.(2002) ]{Ghil} Ghil, M., Allen, R. M., Dettinger,
M. D., Ide, K., Kondrashov, D., Mann, M. E., Robertson, A., Saunders,
A., Tian, Y., Varadi, F., Yiou, P., 2002. Advanced spectral methods
for climatic time series. Reviews of Geophysics 40, 3.1\textendash 3.41
(SSA-MTM tool kit for spectral analysis).

\bibitem[Press et al.(1997) ]{Press}Press,W. H., Teukolsky, S. A.,
Vetterling, W. T., and Flannery, B. P.: Numerical Recipes in C, 2nd
Edn., Cambridge University Press, 1997.

\bibitem[Parker et al.(1992)]{Parker}Parker, D.E., T.P. Legg, and
C.K. Folland. 1992. A new daily Central England Temperature Series,
1772-1991. Int. J. Clim. 12, 317-342.

\bibitem[Scafetta(2014)]{Scafetta2014}N. Scafetta: The complex planetary
synchronization structure of the solar system. Pattern Recognition
in Physics 2 (2014) 1-19. DOI: 10.5194/prp-2-1-2014.

\bibitem[Holm(2014)]{Holm} Holm, S., 2014. On the alleged coherence
between the global temperature and the Sun\textquoteright s movement.
J. Atmos. Solar-Terrestr. Phys. 110\textendash 111, 23\textendash 27. \end{thebibliography}
\end{document}